\documentclass[%
 reprint,
 superscriptaddress,
 amsmath,amssymb,
 aps,
 prl,
]{revtex4-1}

\usepackage{graphicx}
\usepackage{dcolumn}
\usepackage{bm}

\renewcommand{\textfloatsep}{0.3cm}

\begin{document}

\newlength{\len}
\setlength{\len}{89mm}

\title{Measurement of the cosmic-ray antiproton spectrum at solar minimum 
       with a long-duration balloon flight over Antarctica}

\author{ K.\thinspace Abe }
\affiliation{ Kobe University, Kobe, Hyogo 657-8501, Japan }
\author{ H.\thinspace Fuke }
\affiliation{ Institute of Space and Astronautical Science, Japan Aerospace Exploration Agency (ISAS/JAXA), Sagamihara, Kanagawa 229-8510, Japan }
\author{ S.\thinspace Haino }
\affiliation{ High Energy Accelerator Research Organization (KEK), Tsukuba, Ibaraki 305-0801, Japan }
\author{ T.\thinspace Hams }
\affiliation{ NASA-Goddard Space Flight Center (NASA-GSFC), Greenbelt,MD 20771, USA }
\author{ M.\thinspace Hasegawa }
\affiliation{ High Energy Accelerator Research Organization (KEK), Tsukuba, Ibaraki 305-0801, Japan }
\author{ A.\thinspace Horikoshi }
\affiliation{ High Energy Accelerator Research Organization (KEK), Tsukuba, Ibaraki 305-0801, Japan }
\author{ K.\thinspace C.\thinspace Kim }
\affiliation{ IPST, University of Maryland, College Park, MD 20742, USA }
\author{ A.\thinspace Kusumoto }
\affiliation{ Kobe University, Kobe, Hyogo 657-8501, Japan }
\author{ M.\thinspace H.\thinspace Lee }
\affiliation{ IPST, University of Maryland, College Park, MD 20742, USA }
\author{ Y.\thinspace Makida }
\affiliation{ High Energy Accelerator Research Organization (KEK), Tsukuba, Ibaraki 305-0801, Japan }
\author{ S.\thinspace Matsuda }
\affiliation{ High Energy Accelerator Research Organization (KEK), Tsukuba, Ibaraki 305-0801, Japan }
\author{ Y.\thinspace Matsukawa }
\affiliation{ Kobe University, Kobe, Hyogo 657-8501, Japan }
\author{ J.\thinspace W.\thinspace Mitchell }
\affiliation{ NASA-Goddard Space Flight Center (NASA-GSFC), Greenbelt,MD 20771, USA }
\author{ J.\thinspace Nishimura }
\affiliation{ The University of Tokyo, Bunkyo, Tokyo 113-0033, Japan }
\author{ M.\thinspace Nozaki }
\affiliation{ High Energy Accelerator Research Organization (KEK), Tsukuba, Ibaraki 305-0801, Japan }
\author{ R.\thinspace Orito }
\affiliation{ Kobe University, Kobe, Hyogo 657-8501, Japan }
\author{ J.\thinspace F.\thinspace Ormes }
\affiliation{ University of Denver, Denver, CO 80208, USA }
\author{ K.\thinspace Sakai }
\thanks{Corresponding author}
\altaffiliation{Present address: NASA-Goddard Space Flight Center}
\email{Kenichi.Sakai@nasa.gov}
\affiliation{ The University of Tokyo, Bunkyo, Tokyo 113-0033, Japan }
\author{ M.\thinspace Sasaki }
\affiliation{ NASA-Goddard Space Flight Center (NASA-GSFC), Greenbelt,MD 20771, USA }
\author{ E.\thinspace S.\thinspace Seo }
\affiliation{ IPST, University of Maryland, College Park, MD 20742, USA }
\author{ R.\thinspace Shinoda }
\affiliation{ The University of Tokyo, Bunkyo, Tokyo 113-0033, Japan }
\author{ R.\thinspace E.\thinspace Streitmatter }
\affiliation{ NASA-Goddard Space Flight Center (NASA-GSFC), Greenbelt,MD 20771, USA }
\author{ J.\thinspace Suzuki }
\affiliation{ High Energy Accelerator Research Organization (KEK), Tsukuba, Ibaraki 305-0801, Japan }
\author{ K.\thinspace Tanaka }
\affiliation{ High Energy Accelerator Research Organization (KEK), Tsukuba, Ibaraki 305-0801, Japan }
\author{ N.\thinspace Thakur }
\affiliation{ University of Denver, Denver, CO 80208, USA }
\author{ T.\thinspace Yamagami }
\affiliation{ Institute of Space and Astronautical Science, Japan Aerospace Exploration Agency (ISAS/JAXA), Sagamihara, Kanagawa 229-8510, Japan }
\author{ A.\thinspace Yamamoto }
\affiliation{ High Energy Accelerator Research Organization (KEK), Tsukuba, Ibaraki 305-0801, Japan }
\affiliation{ The University of Tokyo, Bunkyo, Tokyo 113-0033, Japan }
\author{ T.\thinspace Yoshida }
\affiliation{ Institute of Space and Astronautical Science, Japan Aerospace Exploration Agency (ISAS/JAXA), Sagamihara, Kanagawa 229-8510, Japan }
\author{ K.\thinspace Yoshimura }
\affiliation{ High Energy Accelerator Research Organization (KEK), Tsukuba, Ibaraki 305-0801, Japan }

\date{\today}

\begin{abstract}
The energy spectrum of cosmic-ray antiprotons ($\bar{p}$'s) from 0.17 to 3.5 GeV has been measured using 7886 $\bar{p}$'s 
detected by BESS-Polar II during a long-duration flight over Antarctica near solar minimum in December 2007 and January 2008.
This shows good consistency with secondary $\bar{p}$ calculations.
Cosmologically primary $\bar{p}$'s have been investigated by comparing measured and calculated $\bar{p}$ spectra.
BESS-Polar II data show no evidence of primary $\bar{p}$'s from evaporation of primordial black holes.
\end{abstract}

\pacs{Valid PACS appear here}
\maketitle

Precise measurement of the cosmic-ray antiproton ($\bar{p}$) spectrum is crucial
to investigations of conditions in the early universe and cosmic-ray propagation.
Most cosmic-ray $\bar{p}$'s are produced by interactions of cosmic-ray nuclei with the interstellar gas.
The energy spectrum of these ``secondary'' $\bar{p}$'s peaks near 2 GeV, decreasing sharply below and above 
due to the kinematics of $\bar{p}$ production and to the local interstellar (LIS) proton spectrum.
The secondary $\bar{p}$'s offer a unique probe \cite{GA92,YO01,ST04} of cosmic-ray propagation and solar modulation.
Cosmologically ``primary'' sources have also been suggested, including the annihilation of dark-matter particles and the evaporation of
primordial black holes (PBH)  by Hawking radiation \cite{HA75}.

Small PBHs, formed in the early Universe by initial density fluctuations, phase
transitions, or the collapse of cosmic strings, might have a significant evaporation rate at
the current age of the Universe and could contribute to the measured $\bar{p}$ spectrum
at low energies \cite{MA96}.
Because the predicted LIS PBH $\bar{p}$ spectrum peaks at $\sim150$ MeV, this would be
strongly influenced by solar modulation, so a search is most sensitive at solar minimum \cite{MI96}.

BESS95+97 showed that the $\bar{p}$ spectrum peaks around 2 GeV \cite{OR00},
and measurements by BESS and other experiments have shown that $\bar{p}$'s are predominantly secondary \cite{MI04}.
However, the low-energy $\bar{p}$ spectrum measured by BESS95+97 at the previous solar minimum was slightly 
flatter than predicted by secondary models.
Although this suggested the possible presence of primary $\bar{p}$'s,
the large statistical error of the BESS95+97 data did not allow a firm conclusion.
BESS-Polar \cite{YA02,MI04,YO04,AJ00,HA04,HA04_2} was developed to evaluate the possibility
of excess low-energy $\bar{p}$ flux, with unprecedented precision,
using long-duration solar-minimum flights over Antarctica. 
BESS-Polar I flew in December 2004 \cite{YA08,YO08,SA08,AB08}
and BESS-Polar II \cite{YO08} flew near solar minimum in December 2007 and January 2008.
Here, we report measurements of cosmic-ray $\bar{p}$'s from 0.17 GeV to 3.5 GeV by BESS-Polar II
and discuss the implications for secondary models and possible primary sources.

\begin{table*}
\caption{\label{tab:table1}$\bar{p}$ flux at the top of atmosphere
                with statistical (first) and systematic (second) errors.
                $N_{\bar{\rm p}}$ and $N_{BG}$ are the number of observed
                $\bar{p}$'s and estimated background events.
                The mean energy for each range was calculated using the measured $\bar{p}$ energies.}
\begin{ruledtabular}
\begin{tabular}{ccccc|ccccc}
\multicolumn{2}{c}{
  \begin{tabular}{@{}cc@{}}
    \multicolumn{2}{c}{Kinetic energy (GeV)}\\ \hline
    \hspace{0.3cm}range\hspace{0.3cm} & mean
  \end{tabular}
} &
\begin{tabular}{@{}c@{}} $N_{\bar{\rm p}}$ \end{tabular} &
\begin{tabular}{@{}c@{}} $N_{BG}$ \end{tabular} &
\begin{tabular}{@{}c@{}}
  ${\bar{\rm p}}$ flux\\
  (m$^{-2}$sr$^{-1}$s$^{-1}$GeV$^{-1}$)
\end{tabular} &
\multicolumn{2}{c}{
  \begin{tabular}{@{}cc@{}}
    \multicolumn{2}{c}{Kinetic energy (GeV)}\\ \hline
    \hspace{0.3cm}range\hspace{0.3cm} & mean
  \end{tabular}
} &
\begin{tabular}{@{}c@{}} $N_{\bar{\rm p}}$ \end{tabular} &
\begin{tabular}{@{}c@{}} $N_{BG}$ \end{tabular} &
\begin{tabular}{@{}c@{}}
  ${\bar{\rm p}}$ flux\\
  (m$^{-2}$sr$^{-1}$s$^{-1}$GeV$^{-1}$)
\end{tabular}\\ \hline
\rule{0mm}{3.5mm}0.17--0.23 & 0.20 &  29 &  0.0 & 3.56$^{\raisebox{0ex}[0ex][-1ex]{\tiny{$+$0.88$+$0.42}}}_{\raisebox{0.1ex}{\tiny{$-$0.78$-$0.42}}}\times 10^{-3}$ & 
0.98--1.07 & 1.03 & 238 &  0.1 & 1.75$^{\raisebox{0ex}[0ex][-1ex]{\tiny{$+$0.15$+$0.13}}}_{\raisebox{0.1ex}{\tiny{$-$0.15$-$0.13}}}\times 10^{-2}$ \\
0.23--0.27 & 0.25 &  26 &  0.0 & 4.53$^{\raisebox{0ex}[0ex][-1ex]{\tiny{$+$1.23$+$0.53}}}_{\raisebox{0.1ex}{\tiny{$-$1.10$-$0.53}}}\times 10^{-3}$ & 
1.07--1.17 & 1.12 & 283 &  0.2 & 1.91$^{\raisebox{0ex}[0ex][-1ex]{\tiny{$+$0.15$+$0.15}}}_{\raisebox{0.1ex}{\tiny{$-$0.15$-$0.15}}}\times 10^{-2}$ \\
0.27--0.32 & 0.30 &  38 &  0.0 & 5.09$^{\raisebox{0ex}[0ex][-1ex]{\tiny{$+$1.13$+$0.50}}}_{\raisebox{0.1ex}{\tiny{$-$1.03$-$0.50}}}\times 10^{-3}$ & 
1.17--1.28 & 1.23 & 304 &  0.6 & 1.82$^{\raisebox{0ex}[0ex][-1ex]{\tiny{$+$0.14$+$0.14}}}_{\raisebox{0.1ex}{\tiny{$-$0.14$-$0.14}}}\times 10^{-2}$ \\
0.32--0.37 & 0.35 &  69 &  0.0 & 7.55$^{\raisebox{0ex}[0ex][-1ex]{\tiny{$+$1.16$+$0.43}}}_{\raisebox{0.1ex}{\tiny{$-$1.07$-$0.43}}}\times 10^{-3}$ & 
1.28--1.40 & 1.34 & 399 &  1.7 & 2.28$^{\raisebox{0ex}[0ex][-1ex]{\tiny{$+$0.15$+$0.17}}}_{\raisebox{0.1ex}{\tiny{$-$0.15$-$0.17}}}\times 10^{-2}$ \\
0.37--0.41 & 0.39 &  44 &  0.0 & 8.05$^{\raisebox{0ex}[0ex][-1ex]{\tiny{$+$1.63$+$0.39}}}_{\raisebox{0.1ex}{\tiny{$-$1.49$-$0.39}}}\times 10^{-3}$ & 
1.40--1.53 & 1.47 & 412 &  3.5 & 2.07$^{\raisebox{0ex}[0ex][-1ex]{\tiny{$+$0.14$+$0.16}}}_{\raisebox{0.1ex}{\tiny{$-$0.14$-$0.16}}}\times 10^{-2}$ \\
0.41--0.44 & 0.42 &  56 &  0.0 & 9.19$^{\raisebox{0ex}[0ex][-1ex]{\tiny{$+$1.65$+$0.45}}}_{\raisebox{0.1ex}{\tiny{$-$1.42$-$0.45}}}\times 10^{-3}$ & 
1.53--1.68 & 1.60 & 466 &  6.2 & 2.10$^{\raisebox{0ex}[0ex][-1ex]{\tiny{$+$0.14$+$0.17}}}_{\raisebox{0.1ex}{\tiny{$-$0.14$-$0.17}}}\times 10^{-2}$ \\
0.44--0.48 & 0.46 &  68 &  0.0 & 9.95$^{\raisebox{0ex}[0ex][-1ex]{\tiny{$+$1.58$+$0.51}}}_{\raisebox{0.1ex}{\tiny{$-$1.46$-$0.51}}}\times 10^{-3}$ & 
1.68--1.84 & 1.75 & 485 &  9.0 & 1.91$^{\raisebox{0ex}[0ex][-1ex]{\tiny{$+$0.13$+$0.16}}}_{\raisebox{0.1ex}{\tiny{$-$0.13$-$0.16}}}\times 10^{-2}$ \\
0.48--0.53 & 0.50 &  87 &  0.0 & 1.14$^{\raisebox{0ex}[0ex][-1ex]{\tiny{$+$0.16$+$0.06}}}_{\raisebox{0.1ex}{\tiny{$-$0.15$-$0.06}}}\times 10^{-2}$ & 
1.84--2.01 & 1.92 & 555 & 11.5 & 2.05$^{\raisebox{0ex}[0ex][-1ex]{\tiny{$+$0.13$+$0.17}}}_{\raisebox{0.1ex}{\tiny{$-$0.12$-$0.17}}}\times 10^{-2}$ \\
0.53--0.57 & 0.55 &  84 &  0.0 & 9.30$^{\raisebox{0ex}[0ex][-1ex]{\tiny{$+$1.41$+$0.52}}}_{\raisebox{0.1ex}{\tiny{$-$1.32$-$0.52}}}\times 10^{-3}$ & 
2.01--2.20 & 2.11 & 632 & 12.9 & 2.18$^{\raisebox{0ex}[0ex][-1ex]{\tiny{$+$0.12$+$0.17}}}_{\raisebox{0.1ex}{\tiny{$-$0.12$-$0.17}}}\times 10^{-2}$ \\
0.57--0.63 & 0.60 & 122 &  0.0 & 1.26$^{\raisebox{0ex}[0ex][-1ex]{\tiny{$+$0.15$+$0.07}}}_{\raisebox{0.1ex}{\tiny{$-$0.14$-$0.07}}}\times 10^{-2}$ & 
2.20--2.41 & 2.31 & 622 & 13.7 & 1.88$^{\raisebox{0ex}[0ex][-1ex]{\tiny{$+$0.11$+$0.16}}}_{\raisebox{0.1ex}{\tiny{$-$0.11$-$0.16}}}\times 10^{-2}$ \\
0.63--0.68 & 0.65 & 131 &  0.0 & 1.20$^{\raisebox{0ex}[0ex][-1ex]{\tiny{$+$0.14$+$0.07}}}_{\raisebox{0.1ex}{\tiny{$-$0.13$-$0.07}}}\times 10^{-2}$ & 
2.41--2.64 & 2.53 & 678 & 13.8 & 1.95$^{\raisebox{0ex}[0ex][-1ex]{\tiny{$+$0.11$+$0.16}}}_{\raisebox{0.1ex}{\tiny{$-$0.11$-$0.16}}}\times 10^{-2}$ \\
0.68--0.75 & 0.71 & 154 &  0.0 & 1.32$^{\raisebox{0ex}[0ex][-1ex]{\tiny{$+$0.14$+$0.08}}}_{\raisebox{0.1ex}{\tiny{$-$0.14$-$0.08}}}\times 10^{-2}$ & 
2.64--2.89 & 2.76 & 637 & 13.3 & 1.77$^{\raisebox{0ex}[0ex][-1ex]{\tiny{$+$0.10$+$0.15}}}_{\raisebox{0.1ex}{\tiny{$-$0.10$-$0.15}}}\times 10^{-2}$ \\
0.75--0.82 & 0.78 & 157 &  0.0 & 1.30$^{\raisebox{0ex}[0ex][-1ex]{\tiny{$+$0.15$+$0.08}}}_{\raisebox{0.1ex}{\tiny{$-$0.14$-$0.08}}}\times 10^{-2}$ & 
2.89--3.16 & 3.00 & 494 & 12.5 & 1.90$^{\raisebox{0ex}[0ex][-1ex]{\tiny{$+$0.12$+$0.22}}}_{\raisebox{0.1ex}{\tiny{$-$0.12$-$0.23}}}\times 10^{-2}$ \\
0.82--0.89 & 0.86 & 209 &  0.0 & 1.84$^{\raisebox{0ex}[0ex][-1ex]{\tiny{$+$0.17$+$0.11}}}_{\raisebox{0.1ex}{\tiny{$-$0.16$-$0.11}}}\times 10^{-2}$ & 
3.16--3.46 & 3.28 & 213 & 11.5 & 1.64$^{\raisebox{0ex}[0ex][-1ex]{\tiny{$+$0.18$+$0.22}}}_{\raisebox{0.1ex}{\tiny{$-$0.17$-$0.22}}}\times 10^{-2}$ \\
0.89--0.98 & 0.94 & 194 &  0.0 & 1.51$^{\raisebox{0ex}[0ex][-1ex]{\tiny{$+$0.15$+$0.10}}}_{\raisebox{0.1ex}{\tiny{$-$0.14$-$0.10}}}\times 10^{-2}$ \\
\end{tabular}
\end{ruledtabular}
\end{table*}

BESS-Polar is a high-resolution magnetic-rigidity spectrometer. 
A uniform field of 0.8 T is produced in a thin superconducting solenoid filled with drift-chamber tracking detectors.
Particle trajectories are determined by fitting up to 52 hit points with a resolution of $\sim140 \mu$m in the bending plane, 
giving a magnetic-rigidity ($\equiv Pc/Ze$) resolution of $0.4\%$ at 1 GV and an overall maximum detectable rigidity (MDR) of 240 GV.
Upper (UTOF) and lower (LTOF) scintillator hodoscopes measure time-of-flight (TOF)
and d$E$/d$x$ and provide the event trigger. For BESS-Polar II $\bar{p}$ measurements, the acceptance is 0.23 m$^{2}$sr.
TOF resolution between the UTOF and LTOF is 120 ps, giving a $\beta^{-1}$ resolution of 2.5\%.
A threshold-type Cherenkov counter (ACC), using a silica aerogel
radiator with optical index $n = 1.03$, rejects $e^{-}$ and $\mu^{-}$
backgrounds by a factor of 6100 to identify $\bar{p}$'s up to 3.5 GeV \cite{YA98}.
A thin scintillator middle-TOF (MTOF) on the
lower surface of the solenoid bore detects low-energy particles
that cannot penetrate the magnet wall.
TOF resolution between the UTOF and MTOF is 320 ps.
In the present analysis, the MTOF was used to independently verify the procedure for eliminating interacting 
upward-going protons that could mimic low-energy $\bar{p}$'s.

\begin{figure}[b]
  \begin{center}
    \includegraphics[width=\len]{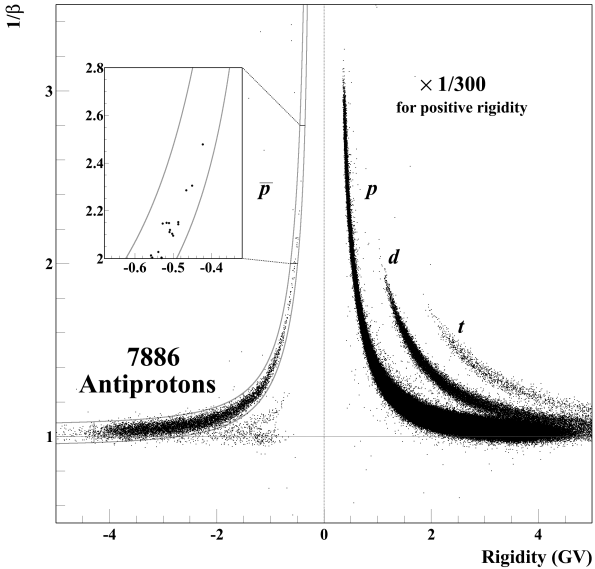}
    \caption {The $\beta^{-1}$ versus rigidity plot and $\bar{p}$ mass selection band after d$E$/d$x$ and ACC cuts. 
                     For clarity, only 1 in 300 positive-rigidity events is shown so only a few $e^{+}$ or $\mu^{+}$ can be seen.
                    The lowest energy $\bar{p}$'s are shown in the inset figure.}
    \label{Fig:figure1}
  \end{center}
\end{figure}

BESS-Polar II was launched on December 23, 2007, from Williams Field, near the US McMurdo
Station in Antarctica, observing for 24.5 days with the magnet energized.
The float altitude was 34 km to 38 km (residual air of 5.8 g/cm$^{2}$ on average), and the cutoff rigidity was below 0.5 GV.
$4.7 \times 10^9$ events were acquired with no inflight event selection as 13.6 terabytes of data.

In flight, most detectors and instrument systems operated well, with expected performance. 
Although the central tracker exhibited high-voltage fluctuations,
normal resolution was preserved for more than 90\% of the observation time by using algorithms that 
calibrate the tracker over short time intervals and depend on its high-voltage state.
Two TOF PMTs with high-voltage control problems were turned off, one on a UTOF paddle (of 10) and one on an LTOF paddle (of 12).
Requiring two good PMTs on each paddle reduced acceptance $\sim20\%$.

Analysis was performed  as described in Ref. \cite{AB08}.
The same selection criteria were applied for $\bar{p}$'s and protons because they behave similarly
in the symmetric configuration of BESS-Polar, except for deflection direction.

Figure~\ref{Fig:figure1} shows $\beta^{-1}$ versus rigidity plots for events surviving d$E$/d$x$ and ACC cuts.
A clean, narrow band of 7886 $\bar{p}$'s mirrors the protons.
The calculated $e^{-}$ and $\mu^{-}$ background is 0.0\%, 1.0\%, and 2.3\% in the 0.2--1.0 GeV, 1.0--2.0 GeV, and 2.0--3.5 GeV energy bands.
Other backgrounds, such as albedo, mismeasured positive-rigidity particles, and re-entrant albedo, were negligible.

The differential flux of $\bar{p}$'s at the top of atmosphere
($\Phi_{\rm TOA}$) integrated over d$E$ can be expressed as:
\begin{eqnarray}
\Phi_{\rm TOA} {\rm d}E = (N_{\rm TOI} - N_{\rm atmos})/\varepsilon_{\rm air}/(S \Omega \cdot T_{\rm live})\\
\label{eq:flux1}
N_{\rm TOI} = (N_{\rm \bar{p}} - N_{\rm BG})/(\varepsilon_{\rm det} \cdot \varepsilon_{\rm non-int})
\end{eqnarray}
where $T_{\rm live}$ is live time, and $N_{\rm \bar{p}}$ and $N_{\rm BG}$ are numbers of observed $\bar{p}$ candidates and
expected background particles.
For the present analysis $T_{\rm live} =$ 1286460 seconds.
The effective geometric acceptance, including noninteraction efficiency $(S\Omega \cdot \varepsilon_{\rm non-int})$, 
was calculated using GEANT3 as 0.133 $\pm$ 0.011 m$^{2}$sr at 0.2 GeV and 0.159 $\pm$ 0.008 m$^{2}$sr at 2.0 GeV,
with errors estimated from differences relative to GEANT4.
The detection efficiency for $\bar{p}$'s ($\varepsilon_{\rm det}$) was calculated using a noninteracting proton sample
as 81.4 $\pm$ 0.1 \% at 0.2 GeV and 60.0 $\pm$ 0.2 \% at 2.0 GeV.
To obtain $\Phi_{\rm TOA}$, 
corrections were applied for $\bar{p}$ survival probability \cite{ST99} in the residual atmosphere ($\varepsilon_{\rm air}$)
and estimated atmospheric $\bar{p}$ production ($N_{\rm atmos}$).
$\varepsilon_{\rm air}$ was estimated as 85.6 $\pm$ 2.0 \% at 0.2 GeV and 89.8 $\pm$ 2.0 \% at 2.0 GeV.
$N_{\rm atmos}$, 17.6 $\pm$ 3.2 \% of the detected $\bar{p}$'s at 0.2 GeV and 27.6 $\pm$ 5.0 \% at 2.0 GeV, 
was calculated by solving simultaneous transport equations \cite{ST99} with 
adjusted interaction length ($\lambda$) and tertiary production \cite{YA05}.
The uncertainty in this calculation is 18.1\% $(=(5.0\%^2({\rm air ~ depth})+8.9\%^2(\lambda)+15.0\%^2({\rm tertiary}))^{1/2})$.

\begin{figure}[t]
  \begin{center}
    \includegraphics[width=\len]{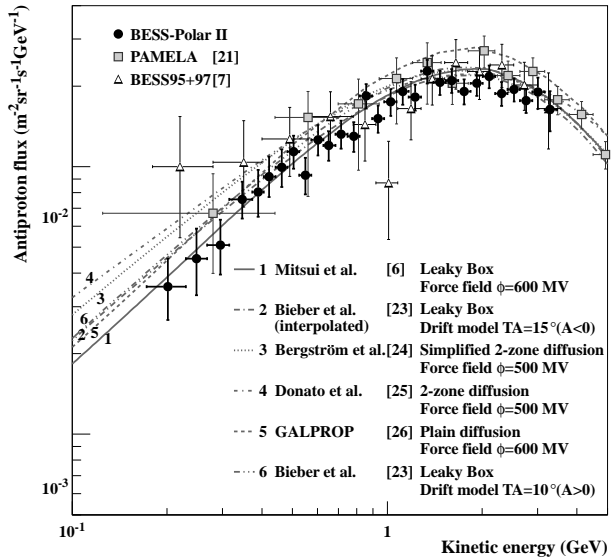}
    \caption {Solar minimum BESS-Polar II, BESS95+97 and PAMELA TOA $\bar{p}$ fluxes and secondary model calculations..}
    \label{Fig:figure2}
  \end{center}
\end{figure}

Table~\ref{tab:table1} gives the flux of $\bar{p}$'s at TOA 
from 0.17 to 3.5 GeV with the statistical (first) and systematic (second) errors.
The dominant systematics are atmospheric subtraction and detection efficiency.
A rapid change in efficiency due to the ACC veto increases the systematic uncertainty in the two highest bins.

Fig.~\ref{Fig:figure2} shows the BESS-Polar II $\bar{p}$ spectrum 
with BESS95+97 and PAMELA \cite{AD10} measurements and solar-minimum 
secondary calculations \cite{MI96, MID,BI99C, BE99,DO01,ST98}. 
Curve 1 uses Mitsui et al. \cite{MI96, MID} data with force-field modulation of 600 MV from
the best fit to the BESS-Polar II proton spectrum.
Curve 2 was generated by interpolating model calculations supplied by Bieber et al. \cite{BI99C}
for negative solar magnetic field polarity (A$<$0). 
The tilt angle of 15$^\circ$(A$<$0) is the best fit to the BESS-Polar II proton data. 
Curve 6 is the published A$>$0 solar-minimum calculation \cite{BI99C} for comparison to the BESS95+97 measurements. 
Curves 3 \cite{BE99} and 4 \cite{DO01} are also published solar-minimum calculations. 
Curve 5 was generated using the GALPROP model \cite{ST98} with 600 MV force-field modulation.
Improved statistical precision of the measured $\bar{p}$ flux results from 14 and 30 times 
more events below 1 GeV than BESS95+97 and PAMELA, respectively.
The BESS-Polar II and PAMELA spectra generally agree in shape, but differ in absolute flux. 
The weighted mean difference, with combined uncertainties, is 14 $\pm$ 5\%, calculated near 2 GeV to reduce modulation effects.
Both are consistent with solar-minimum secondary calculations.
Neither exhibits the flattening at low energies found by BESS95+97, although the differences are statistically small.

\begin{figure}[t]
  \begin{center}
    \includegraphics[width=\len]{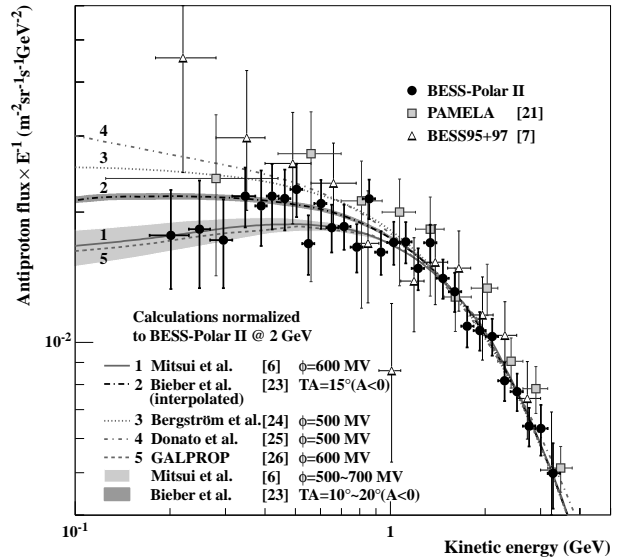}
    \caption {Comparison of $\bar{p}$ flux shapes with secondary calculations normalized to BESS-Polar II flux at  2 GeV.
              Sensitivity to uncertainty in the force-field modulation parameter is shown by the lower shaded band. 
              The small sensitivity of secondary $\bar{p}$ drift calculations at solar minimum to tilt angle is illustrated by the upper shaded band.}
    \label{Fig:figure3}
  \end{center}
\end{figure}

The evident differences among the calculations shown in Fig.~\ref{Fig:figure2} arise from several factors 
that can affect the normalization or shape of the spectrum:  
(1) definition of the primary proton and helium spectra,
(2) incomplete knowledge of nuclear physics in propagation, 
(3) parameters and models of propagation in the Galaxy, 
and (4) modulation in the heliosphere. 
Variation in the absolute fluxes of interstellar protons and helium, 
for instance, affects the absolute flux of $\bar{p}$'s, but not the spectral shape.

Precise measurement of the low-energy $\bar{p}$ spectrum by BESS-Polar II allows secondary flux calculations 
to be evaluated by comparing observed and predicted spectral shapes, as shown in Fig.~\ref{Fig:figure3}.
The calculations are normalized to BESS-Polar II at 2 GeV
to focus on their shapes. 
The calculated spectra and data points are also multiplied by $E_{k}^{-1}$ to emphasize differences at low energies. 
\textit{The observed data are not normalized.}
Chi-square ($\chi^2$) calculated with BESS-Polar II data and the normalized secondary $\bar{p}$ calculations in Fig.~\ref{Fig:figure3} 
are 0.61 (1), 0.61 (2), 1.32 (3), 1.70 (4), 0.67 (5).
The shape variation from uncertainty in the level of solar modulation is illustrated by the lower shaded band, 
calculated with the Mitsui et al. model \cite{MI96,MID} and modulation parameters of 500 MV ($\chi^2$=0.81) and 700 MV ($\chi^2$=0.52).
The small sensitivity of drift calculations to tilt angle is shown by the upper shaded band using A$<$0 Bieber et al. data \cite{BI99C} at 10$^\circ$
and 20$^\circ$ (interpolated). 
In both cases, the change in spectral shape is small compared to differences arising from propagation models, 
because of the peaked shape of the LIS secondary $\bar{p}$ spectrum.
BESS-Polar II results are more consistent with models (curves 1, 2, and 5) without low-energy $\bar{p}$'s 
from tertiary interactions (curve 3) or a soft spectrum from diffusive reacceleration (curve 4).

\begin{figure}[t]
  \begin{center}
    \includegraphics[width=\len]{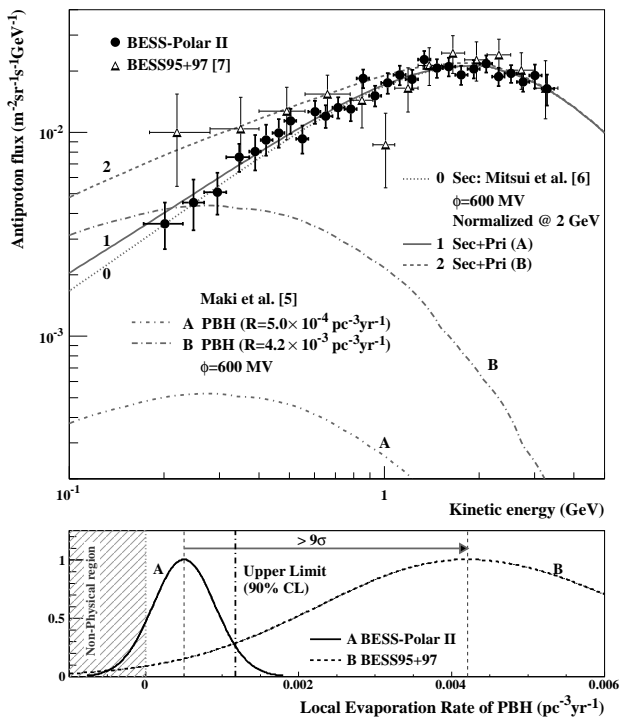}
    \caption {(Top) Possible primary $\bar{p}$ fluxes from PBH evaporation calculated for
              BESS-Polar II (A) and BESS95+97 (B) by fitting differences of the measured spectra from the Mitsui secondary $\bar{p}$ spectrum.
              (Bottom) PBH evaporation rate ($\cal R$) distributions. 
              Values of $\cal R <$ 0 are non-physical.}
    \label{Fig:figure4}
  \end{center}
\end{figure}

The likelihood of primary $\bar{p}$'s from PBH evaporation 
can be quantified by a model-dependent evaporation rate ($\cal R$) determined by fitting a PBH model spectrum to the difference 
of a secondary calculation from the measured flux. 
$\cal R$ is positive (physical) only if the measured flux exceeds the secondary prediction. 
To avoid bias from uncertainties in the predicted absolute flux,
the secondary calculation is normalized to the measurements at the spectral peak (2 GeV) as in Fig.~\ref{Fig:figure3}.
Comparing the models shown in Fig.~\ref{Fig:figure3}  to the measurements, only 1 and 5 give a significant, and almost identical, excess. 
We use curve 1, a slightly better fit to the measured spectrum, to calculate $\cal R$.
Using the Maki et al. PBH model \cite{MA96} with force-field modulation gives 
${\cal R} = 5.0^{\raisebox{0ex}[0ex][-1ex]{\scriptsize{$+4.1$}}}_{\raisebox{0.1ex}{\scriptsize{$-4.0$}}} \times 10^{-4} \rm{pc}^{-3}\rm{yr}^{-1}$,
as shown in Fig.~\ref{Fig:figure4}.
This excludes by more than 9 sigma the slight possibility of primary $\bar{p}$'s suggested 
by ${\cal R} = 4.2^{\raisebox{0ex}[0ex][-1ex]{\scriptsize{$+1.8$}}}_{\raisebox{0.1ex}{\scriptsize{$-1.9$}}} \times 10^{-3} \rm{pc}^{-3}\rm{yr}^{-1}$
from BESS95+97 data with the same models and modulation.
We also find a 90\% confidence level upper limit of ${\cal R} \sim1.2 \times 10^{-3} \rm{pc}^{-3}\rm{yr}^{-1}$.
This is almost insensitive to modulation
(500 MV: ${\cal R} = 1.0 \times 10^{-3} \rm{pc}^{-3}\rm{yr}^{-1}$, 600 MV: ${\cal R} = 1.2 \times 10^{-3} \rm{pc}^{-3}\rm{yr}^{-1}$, 
700 MV: ${\cal R} = 1.3 \times 10^{-3} \rm{pc}^{-3}\rm{yr}^{-1}$).

The affects of charge-sign dependent modulation in the A$>$0 and A$<$0 solar magnetic field polarities 
on secondary and possible PBH primary $\bar{p}$ fluxes differ considerably 
because of their spectral shapes.
Curves 2 and 6 in Fig.~\ref{Fig:figure2} indicate that the differences in the solar-minimum secondary fluxes are small. 
However, the predicted LIS PBH $\bar{p}$ spectrum peaks near the lower end of the BESS-Polar II energy range, 
and solar polarity would strongly affect the contribution of primaries to the measured low-energy flux. 
The primary $\bar{p}$ flux should be suppressed for A$>$0 and higher for A$<$0. 
Thus, solar polarity cannot explain the excess reported by BESS95+97 or the negligible excess in the BESS-Polar II results reported here.
\textit{Within statistics, the BESS-Polar II data show no evidence of primary $\bar{p}$'s from PBH evaporation.}

The BESS-Polar  collaboration is supported in Japan
by the Grant-in-Aid `KAKENHI' for Specially Promoted and Basic Researches, MEXT-JSPS, and in the U.S. by NASA. 
Balloon flight operations were carried out by the NASA Columbia Scientific Balloon Facility and the National
Science Foundation United States Antarctic Program. We would like to express our sincere thanks for their continuous
professional support.

\bibliography{ref.bib,string_abbrv.bib}

\end{document}